\documentclass[aps,pra,preprint]{revtex4-1}   

\usepackage{graphicx}
\usepackage{amsmath}
\usepackage{amssymb}
\usepackage{mathrsfs}
\usepackage{amsfonts}
\usepackage{dsfont}
\usepackage{dcolumn}
\usepackage{bm}
\usepackage{booktabs}
\usepackage{textcomp}

\everymath{\displaystyle}

\newcommand{\be}{\bm{e}}

\providecommand{\abs}[1]{\lvert#1\rvert}

\providecommand{\ave}[1]{\langle#1\rangle}

\begin{document}

\title{Comment on ``Semiclassical and Quantum Analysis of a Focussing Free Particle Hermite Wavefunction'', by Paul Strange (arXiv:1309.6753 [quant-ph])}
\author{Andrea Aiello$^{1,2}$}
\email{andrea.aiello@mpl.mpg.de}

%
\affiliation{$^1$ Max Planck Institute for the Science of Light, G$\ddot{u}$nther-Scharowsky-Strasse 1/Bau24, 91058 Erlangen,
Germany}
\affiliation{$^2$Institute for Optics, Information and Photonics, University of Erlangen-Nuernberg, Staudtstrasse 7/B2, 91058 Erlangen, Germany}
\date{\today}

\begin{abstract}

In the recent and very enjoyable paper (Paul Strange, \emph{Semiclassical and Quantum Analysis of a Focussing Free Particle Hermite Wavefunction}, arXiv:1309.6753[quant-ph]), Professor Strange has studied a particular solution of the free particle Schr\"{o}dinger equation 
in which the time and space dependence are not separable. After recognizing the fact that ``The Schr\"{o}dinger equation has an identical mathematical form to the
paraxial wave equation [...]'', he claims to ``describe and try to gain insight into an exotic,
apparently accelerating solution of the free particle Schr\"{o}dinger equation that is square integrable
and which also displays some unusual characteristics.'' 
 It is the main aim of this short note to show that the wavefunction described by Prof. Strange is simply one particular  Hermite-Gauss solution of the paraxial wave equation and even more ``exotic'' examples can be found by considering, e.g., Laguerre-Gauss solutions.
\end{abstract}

\maketitle

Professor Strange \cite{Strange} considered a solution of the Schr\"{o}dinger time-dependent equation for
one-dimensional  motion of a free particle of mass $m$,
\begin{align}
i \hbar \frac{\partial \psi(x,t)}{\partial t} = -\frac{\hbar^2}{2 m}\frac{\partial^2 \psi(x,t)}{\partial x^2},
\end{align}
that at $t=0$ takes the form 
\begin{align}
\psi(x,t) = \frac{1}{2^n n!} \left( \frac{m \omega}{\hbar \pi} \right)^{1/4} e^{-m \omega x^2/(2 \hbar)} 
H_n\left( \sqrt{\frac{m \omega}{\hbar}}x\right)
\end{align}
where $\omega = 1/t_c$. This equations is a 1-dimensional case of the more general function 
\begin{align}\label{psiHG}
\psi_{\mu \nu}(x,y,t) = & \; \frac{1/w_0}{\sqrt{2^{\mu + \nu -1} \pi \mu! \nu!}} 
\exp\left(- \frac{x^2 + y^2}{w_0^2}\frac{1}{1 + i t/t_0} \right) \exp\left[- i \left(\mu + \nu +1 \right)\arctan \left( \frac{t}{t_0}\right) \right] \nonumber \\
 & \; \times H_\mu \left( \frac{\sqrt{2} \, x}{w_0 \sqrt{1 + t^2/t_0^2}}\right) \,  H_\nu \left( \frac{\sqrt{2} \, y}{w_0\sqrt{1 + t^2/t_0^2}}\right),
\end{align}
which is a solution of the 2-dimensional   Schr\"{o}dinger equation for
 a free particle of mass $m$
\begin{align}
i \hbar \frac{\partial \psi_{\mu \nu}(x,y,t)}{\partial t} = -\frac{\hbar^2}{2 m}\left[ \frac{\partial^2 \psi_{\mu \nu}(x,y,t)}{\partial x^2} + \frac{\partial^2 \psi_{\mu \nu}(x,y,t)}{\partial y^2}\right],
\end{align}
provided that $t_0 = m w_0^2/(2 \hbar)$, $\mu, \nu \in \{ 0,1,2, \dots\}$ and $w_0>0$ is a free parameter with the dimensions of a length that fixes the size of the wavefunction. The solution \eqref{psiHG} can be easily found by taking a Hermite-Gauss solution $u_{\mu \nu}(x,y,x)$ of the paraxial wave equation \cite{Siegman}
\begin{align}
2 i k \frac{\partial u_{\mu \nu}(x,y,x)}{\partial z} = - \frac{\partial^2  u_{\mu \nu}(x,y,z)}{\partial x^2} - \frac{\partial^2  u_{\mu \nu}(x,y,z)}{\partial y^2},
\end{align}
and making the formal replacements $z \to t$ and $k \to m/\hbar$, where $k$ is the wavenumber of the Hermite-Gauss beam. When $\mu = 0 = \nu$  the wavefunction $\psi_{\mu \nu}(x,y,t)$ reduces to the well-know free-space Gaussian wavepacket 
\begin{align}\label{psiHG2}
\psi_{00}(x,y,t) = & \; \sqrt{ \frac{2}{\pi}} 
\exp\left(- \frac{x^2 + y^2}{w_0^2}\frac{1}{1 + i t/t_0} \right) \exp\left[- i\arctan \left( \frac{t}{t_0}\right) \right],
\end{align}
widely studied in the literature \cite{Darwin}. 

More wavefunctions with counter-intuitive behavior may be found by considering the Laguerre-Gauss solutions of the paraxial wave equation. Consider, for example, the following function:
\begin{align}\label{psiLG}
\psi_{\ell}(x,y,t) = & \; \frac{2^{\frac{1+\ell}{2}}}{\sqrt{ \pi \ell!}}  
\exp\left(- \frac{x^2 + y^2}{w_0^2}\frac{1}{1 + i t/t_0} \right) \frac{ \left( x + i y \right)^{\ell}}{\left[ w_0 \left(1 + i t/t_0 \right) \right]^{1+\ell}} ,
\end{align}
with $\ell \in \{ 0, \pm 1, \pm2, \ldots\}$. Equation \eqref{psiLG} represents an exact, square-integrable solution of the free particle Schr\"{o}dinger equation in 2+1 dimensions, which displays some ``exotic'' features, as orbital angular momentum equal to $\ell \hbar$ without ``rotation''. What does this mean? Well, when thinking about a free particle one usually imagines a little ball propagating along a rectilinear trajectory with constant velocity. However, in quantum mechanics the particle may be initially, at $t=0$, prepared in a state that do not resemble at all a small ball, namely a spatially localized object. Conversely, one can prepare a state where the particle is perfectly delocalized along a ring:
\begin{align}\label{psiLG2}
\abs{\psi_{\ell}(x,y,t=0)}^2 = & \; \frac{2^{1+\ell}}{ \pi \ell!}  
\exp\left(- 2\frac{x^2 + y^2}{w_0^2} \right) \frac{ \left( x^2 + y^2 \right)^{\ell}}{ w_0^{1+\ell}} .
\end{align}
In fact, there is nothing strange about Eq. \eqref{psiLG} from the viewpoint of quantum mechanics. If one calculates the probability density $\rho_\ell =\psi_{\ell}^* \psi_{\ell}$ and the probability current density 
\begin{align}\label{current}
\bm{j}_\ell = \frac{\hbar}{2 m i}\left[ \psi_{\ell}^*\left(\be_x \frac{\partial \psi_{\ell}}{\partial x} + \be_y \frac{\partial \psi_{\ell}}{\partial y}\right) -\left( \be_x \frac{\partial \psi_{\ell}^*}{\partial x} + \be_y \frac{\partial \psi_{\ell}^*}{\partial y}\right) \psi_{\ell}\right] ,
\end{align}
it is not difficult to see that the continuity equation is identically satisfied:
\begin{align}\label{continuity}
\frac{\partial \rho_\ell}{\partial t} + \frac{\partial j_{\ell x}}{\partial x} + \frac{\partial j_{\ell y}}{\partial y} =0 .
\end{align}
As time goes by, the particle described by $\psi_\ell(x,y,t)$ does not move, in the sense that $\ave{x}=0 = \ave{y}$ and $\ave{p_x}=0 = \ave{p_y}$, where $p_x = -i \hbar \partial_x$, etc. The other relevant expectation values are:
\begin{align}\label{ave}
\ave{ x^2 + y^2 } =  \frac{1+\ell}{2} w_0^2 \left( 1 + \frac{t^2}{t_0^2} \right), \qquad \ave{p_x^2 + p_y^2 } = 
\left( 1+ \ell \right) \frac{2 \hbar^2}{w_0^2},
\end{align}
and
\begin{align}\label{aveEn}
\ave{H } =  \frac{ \ave{p_x^2 + p_y^2 }}{2m }  = 
\left( 1+ \ell \right) \frac{ \hbar^2}{mw_0^2}, \qquad \ave{x p_x - y p_x } = \ell \hbar.
\end{align}

These equations describe the motion of a free quantum particle moving in a ring whose radius change with time, as shown in Fig. 1, which illustrates the time evolution of the probability density $\rho_1(x,y,t)$ evaluated for $\ell=1$:
%
%
\begin{figure}[h!]
\begin{center}
\includegraphics[angle=0,width=15truecm]{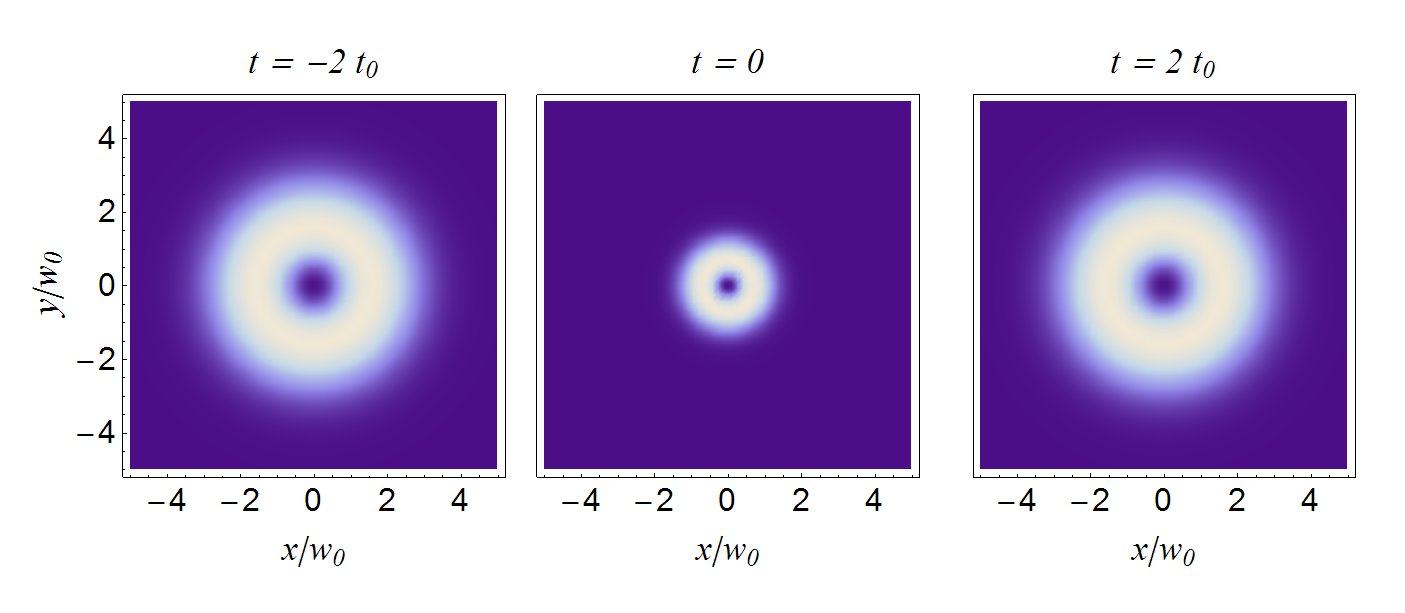}
\caption{\label{fig1} Time evolution of the probability density $\rho_1(x,y,t)$. At $t=-2t_0$ the particle is in the ring at the left column. At $t=0$ it reaches its minimum radius as permitted by the uncertainty principle, and then it begins to expand again.}
\end{center}
\end{figure}
%
%
\newpage
The same ``free-focussing'' phenomenon may be better observed by taking a cross section at $y=0$ of the probability density $\rho_1(x,y,t)$, as shown in Fig. 2.
%
%
\begin{figure}[h!]
\begin{center}
\includegraphics[angle=0,width=7truecm]{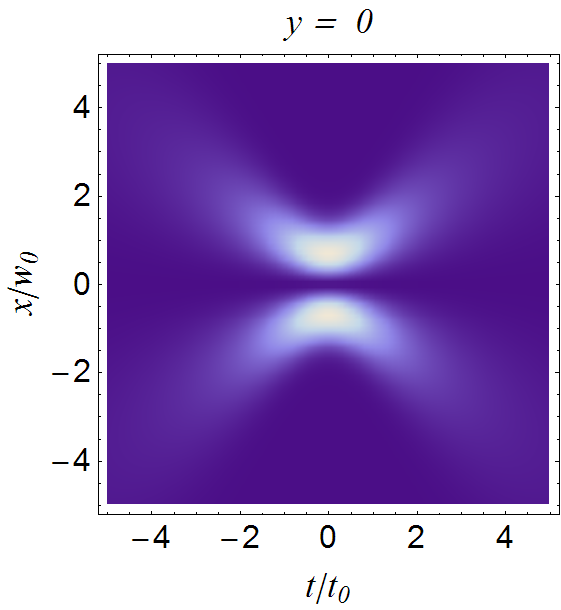}
\caption{\label{fig1} Time evolution of the section of the probability density $\rho_1(x,y=0,t)$ evaluated at $y=0$.}
\end{center}
\end{figure}
%
%
\newpage
Finally, in Fig. 3 we can follow the ``motion'' of the particle by plotting the streamlines of the vector field $\bm{j}_1(x,y,t)$:
%
%
\begin{figure}[h!]
\begin{center}
\includegraphics[angle=0,width=15truecm]{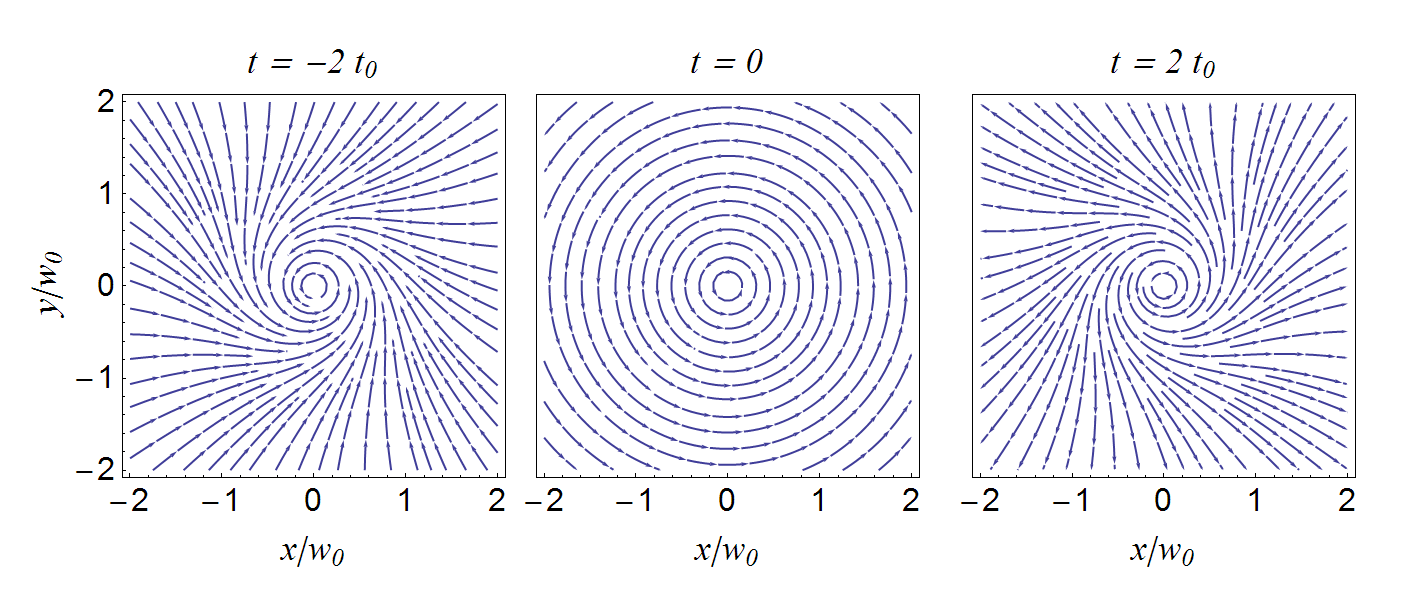}
\caption{\label{fig1} Snapshots of the streamlines plot of the probability current density $\bm{j}_1(x,y,t)$ evaluated for $\ell =1$. Note that the handiness of the rotation does not change at $t=0$.}
\end{center}
\end{figure}
%
%

From the figures above it appears in a clear manner why these phenomena seems counter intuitive if one think of a particle as a bullet. However, from a quantum mechanical point of view the description furnished above is perfectly consistent. 


\begin{thebibliography}{999999}
%
\bibitem{Strange} P. Strange, ``Semiclassical and Quantum Analysis of a Focussing Free Particle Hermite Wavefunction'', arXiv:1309.6753[quant-ph].
%
%
\bibitem{Siegman} A. E. Siegman, \emph{Lasers}, (University Science Books, USA, 1986).
%
\bibitem{Darwin} 
C. G. Darwin,  ``Free motion in quantum mechanics'', Proc. R. Soc. A \textbf{117} 258–93 (1928).
%
%
\end{thebibliography}
\end{document}